\documentclass[11pt]{article}
\usepackage{amsmath}
\usepackage{amsfonts}
\usepackage{amssymb}
\newtheorem{theorem}{Proposition}

\newcommand{\dbar}{\bar{\partial}}
\newcommand{\D}{\mathcal{D}}
\newcommand{\wt}{\widetilde}
\newcommand{\be}{\begin{equation}}
\newcommand{\ee}{\end{equation}}
\newcommand{\bea}{\begin{eqnarray}}
\newcommand{\eea}{\end{eqnarray}}
\newcommand{\beaa}{\begin{eqnarray*}}
\newcommand{\eeaa}{\end{eqnarray*}}

\newcommand{\nn}{\nonumber}
\begin{document}
\title
{On dispersionless BKP hierarchy and its
reductions}
\author{
L.V. Bogdanov\thanks {L.D. Landau ITP, Kosygin str. 2, Moscow
119334, Russia} ~and B.G. Konopelchenko\thanks {Dipartimento di
Fisica dell' Universit\`a di Lecce and Sezione INFN, 73100 Lecce,
Italy}}
\date{}
\maketitle
\begin{abstract} Integrable dispersionless Kadomtsev-Petviashvili (KP)
hierarchy of B type
is considered. Addition formula for the $\tau$-function
and conformally invariant equations
for the dispersionless BKP (dBKP) hierarchy are derived. Symmetry constraints for the dBKP
hierarchy are studied.
\end{abstract}

\begin{center}
\end{center}
\section{Introduction}
Dispersionless integrable hierarchies
play an important role  in the study of different nonlinear phenomena
in various fields of physics and mathematics (see e.g. \cite{KM77}-\cite{17}).
They have attracted recently
a considerable interest (see e.g. \cite{WZ}-\cite{AM1}).

Quasi-classical $\dbar$-dressing method, proposed in \cite{KMR,KM1,KM3,KM2}, gives a new approach
to study the properties of dispersionless integrable hierarchies,
including various addition formulae,
symmetry constraints etc. \cite{BKA,dtau,dconstr}. Most of the results obtained using this approach
are connected with dispersionless Kadomtsev-Pe\-tvi\-a\-shvi\-li (dKP),
modified dispersionless Kadomtsev-Petviashvili (dmKP) and dispersionless
2-dimensional Toda lattice (d2DTL) hierarchies.

In the full `dispersive' case not only the standard KP hierarchy
(the hierarchy of A type), but also
BKP, CKP and DKP hierarchies play an important role \cite{DJKM1,JM,DJKM2}.
In the dispersionless case the study of hierarchies of B type is just
in the very beginning \cite{Takasaki,KM3}.

In the present paper we analyze the dBKP hierarchy in detail.
We formulate the $\dbar$-dressing approach to this hierarchy,
derive an integral formula for the $\tau$-function, obtain
the fundamental equation for the basic homeomorphism which represents
a generating equation for the whole hierarchy. We derive also the
dispersionless addition
formula for the $\tau$-function and obtain conformally invariant equation for
its symmetries. We consider symmetry constraints for the dBKP hierarchy and
find generating equations and Sato functions for the constrained hierarchies.
\section{dKP and dBKP hierarchies}
The quasiclassical $\dbar$-dressing scheme for dispersionless KP hierarchy
\cite{KMR,KM1,KM3,KM2} is based on nonlinear Beltrami equation
\bea
S_{\bar z}=W(z, \bar z, S_z),
\label{ddbar}
\eea
where $\dbar$-data $W$ are localized in the unit disc, $S(z,\bar z,\mathbf{t})=S_0+\wt S$,
$S_0(z,\mathbf{t})=\sum_{n=1}^\infty t_n z^n$, $\wt S$ is analytic outside the unit disc, and
at infinity it has an expansion
$\wt S=\sum_{i=1}^{\infty}\wt S_i(\mathbf{t})z^{-i}$.
The quantity $p=\frac{\partial S}{\partial t_1}$
is a basic homeomorphism \cite{KM1}.
Important role in the theory of dKP hierarchy
is played by the equation
\be
p(z)-p(z_1)+z_1\exp(-D(z_1)S(z))=0,\quad z\in\mathbb{C},
\quad z_1\in\mathbb{C}\setminus D,
\label{DE2}
\ee
(where $D(z)$ is the quasiclassical vertex operator,
$D(z)=\sum_{n=1}^{\infty}\frac{1}{n}\frac{1}{z^n}
\frac{\partial}{\partial t_n}$, $|z| > 1$)
which generates Hamilton-Jacobi equations of the hierarchy
by expansion into the powers of $z^{-1}$ at infinity
(see, e.g., \cite{BKA}). This equation also implies
existence of the $\tau$-function, characterized by the relation
\be
\wt S(z,\mathbf{t})=-D(z)F(\mathbf{t}),
\label{difftauKP}
\ee
and provides the dispersionless addition formula.

Dispersionless BKP hierarchy is characterized by the symmetry condition
\be
S(-z,\mathbf{t})=-S(z,\mathbf{t}),
\label{symB}
\ee
which is preserved only by odd flows of the hierarchy ($t_n$ with odd $n$),
$S_0(z,\mathbf{t})=\sum_{n=0}^\infty t_{2n+1} z^{2n+1}$.
Thus dBKP hierarchy is dKP hierarchy with even times frozen at zero plus symmetry
(\ref{symB}) \cite{Takasaki}.
In terms
of $\dbar$-data this symmetry is provided by the condition
$$
W(-z,S_z)=W(z,S_z).
$$

To obtain the analogue of relation (\ref{DE2}) for dBKP hierarchy,
we introduce B-type quasiclassical vertex operator $\D(z)=
2\sum_{n=0}^{\infty}\frac{z^{-(2n+1)}}{2n+1}
\frac{\partial}{\partial t_{2n+1}}$, $|z| > 1$, characterized by the property
$\D(z_1)S_0(z)=\ln\frac{z_1+z}{z_1-z}$. Then, similar to dKP case \cite{BKA},
we get
\be
\frac{p(z_1)-p(z)}{p(z_1)+p(z)}=\exp(-\D(z_1)S(z)),
\label{DEB}
\ee

This equation generates Hamilton-Jacobi equations of the dBKP hierarchy
by expansion into the powers of $z^{-1}$ at infinity.
The first two Hamilton-Jacobi equations are
\bea
&&
S_y=p^3 + 3up\\
&&
S_t=p^5+5u p^3+vp,
\eea
where $x=t_1$, $y=t_3$, $t=t_5$,
$u=-\partial_x \wt S_1$, $v_x=\frac{5}{3}u_y + 5(u^2)_x$. Compatibility condition
for these equations gives dispersionless BKP equation \cite{KM3}
\be
\textstyle
\frac{1}{5}u_t+u^2u_x-
\frac{1}{3}uu_y-
\frac{1}{3}u_x\partial_x^{-1}u_y-
\frac{1}{9}\partial_x^{-1}u_{yy}=0.
\label{dBKP}
\ee

Relation (\ref{DEB}) for dBKP hierarchy can be also obtained
starting with
(\ref{DE2}). Indeed, using (\ref{DE2}), in the framework of dKP hierarchy
we get a relation
\be
\frac{p(z)-p(z_1)}{p(z)-p(-z_1)}=-\exp(-(D(z_1)-D(-z_1))S(z)),
\label{DEA}
\ee
Then, freezing even times at zero and using symmetry (\ref{symB}),
we get (\ref{DEB})
($\D(z)=D(z)-D(-z)$).

Equation (\ref{DEB}) also implies
existence of the $\tau$-function, characterized by the relation
$$
\wt S(z,\mathbf{t})=-\D(z) F_\text{dBKP}(\mathbf{t}).
$$
Comparing this relation with (\ref{difftauKP}), we come to the conclusion that
$$
2F_\text{dBKP}=F_\text{dKP},
$$
if $F_\text{dKP}$ is taken at zero even times and symmetry condition (\ref{symB})
is satisfied \cite{Takasaki}.
This symmetry condition is equivalent to a simple condition for the
function $F_\text{dKP}$ itself,
namely that its derivatives $\partial F_\text{dKP}/\partial t_{2(n+1)}$
taken at zero even times are equal to zero.

In complete analogy with dKP hierarchy, it is possible to find
explicit representaion of dBKP $\tau$-function as an action for nonlinear
Beltrami equation (\ref{ddbar}) evaluated on its solution.
\begin{theorem}
The function
\bea
F(\mathbf{t})
=-\frac{1}{2\pi\mathrm{i}}\iint_{D}
\left(\frac{1}{2}\wt S_{\bar z}(\mathbf{t}) \wt S_z(\mathbf{t})-
W_{-1}(z,\bar z,S_z(\mathbf{t}))\right)dz\wedge d\bar z,
\label{TAU}
\eea
i.e., the action
for the problem (\ref{ddbar}) evaluated on its solution,
where
$\partial_\eta W_{-1}(z,\bar z,\eta)=W(z,\bar z,\eta)$, and $W$ satisfies a symmetry condition
$$
W(-z,-\bar z,S_z)=W(z,\bar z,S_z),
$$
is a $\tau$-function of
dBKP hierarchy.
\end{theorem}
Variations of the $\dbar$-data define infinitesimal symmetries of the $\tau$-function
\cite{dtau,dconstr}.
One should take into account that these variations should satisfy the symmetry condition.
Considering variations localized in the pair of points $z_0$, $-z_0$, we get a symmetry
\be
\delta F=f(S_z(z_0)),
\label{sym1}
\ee
where $f$ is an arbitrary analytic function. Variations localized on the set of
curves lead to infinitesimal symmetry
\be
\delta F=\sum_{i=1}^N c_i (S_i-\tilde S_i),
\label{sym2}
\ee
where $S_i=S(z_i)$, $\tilde S_i=S(\tilde z_i)$, $z_i$, $\tilde z_i$ are some sets of
points, and $c_i$ are arbitrary constants. Due to the symmetry $S(-z)=-S(z)$ in dBKP case,
it is possible to take $\tilde z_i=-z_i$ and consider symmetries of the form
\be
\delta F=2\sum_{i=1}^N c_i S_i.
\label{sym3}
\ee
\section{Addition formula for dBKP hierarchy}
Expressing $S$ in terms of $F$, from equation (\ref{DEB}) we get
\be
\frac{p(z_1)-p(z_2)}{p(z_1)+p(z_2)}=\frac{z_1-z_2}{z_1+z_2}e^{\D(z_1)\D(z_2)F}.
\label{sys0}
\ee
Using this equation, we obtain a system of linear equations for $p(z_i)$,
\be
p(z_i)(f_{ij}-1)-p(z_j)(f_{ij}+1)=0
\label{sys}
\ee
where
\be
f_{ij}=\frac{z_i-z_j}{z_i+z_j}e^{\D(z_i)\D(z_j)F},\quad 1\leq i,j\leq 3.
\label{f-def}
\ee
To possess nontrivial solutions, this system should have zero determinant,
\beaa
\det
\left(
\begin{array}{ccc}
f_{12}-1&f_{12}+1&0\\
0&f_{23}-1&f_{23}+1\\
f_{13}+1&0&f_{13}-1
\end{array}
\right)
=0.
\eeaa
Thus
\be
(f_{23}+1)(f_{31}+1)(f_{12}+1)=(f_{23}-1)(f_{31}-1)(f_{12}-1),
\label{add1}
\ee
or, equivalently,
\be
f_{23}f_{31}f_{12}+f_{23}+f_{31}+f_{12}=0.
\label{add2}
\ee
This condition gives addition formula for dispersionless BKP hierarchy,
\be
1+
\text{\footnotesize $c_2c_3$}
e^{-(\D_3\D_1+\D_1\D_2)F}+
\text{\footnotesize $c_1c_3$}
e^{-(\D_2\D_3+\D_1\D_2)F}+\text{\footnotesize $c_2c_1$}
e^{-(\D_3\D_1+\D_2\D_3)F}=0,
\label{add3}
\ee
where $\D_i=\D(z_i)$, $c_i=\frac{z_j+z_k}{z_j-z_k}$
($(i,j,k)$ is a cyclic permutation of $(1,2,3)$).
\section{Conformally invariant equations of dBKP hierarchy}
An important object of dispersive integrable hierarchies are discrete
Schwarzian equations, which possess M\"obius symmetry and have a deep connection
with geometry \cite{KS1,KS2}. It was demonstrated in
\cite{dtau}  for KP and 2DTL hierarchies
that dispersionless analogues of these equations are given by
conformally invariant equations of dispersionless hierarchies, which arise also
an a naive continuum limit of discrete Schwarzian equations \cite{KS1}.

In the case of dBKP hierarchy we start with the
evident  analogy of (\ref{sys}), (\ref{add1}) and the formulae in \cite{KS2}
connected with continuum limit of discrete SBKP equation. Using this analogy, we introduce
the function $\Phi$, $p(z_i)=\D_i\Phi$.
To demonstrate existence of $\Phi$, let us rewrite relations (\ref{sys})
in the form
\beaa
\D_2\Phi=\frac{f_{12}-1}{f_{12}+1}\D_1\Phi,\quad
\D_3\Phi=\frac{f_{13}-1}{f_{13}+1}\D_1\Phi
\eeaa
Compatibility of this linear system is implied by addition formula (\ref{add2}).
Thus the function $\Phi$,
$p(z_i)=\D_i\Phi$ exists.

Relations
(\ref{sys0}) imply equation for $\Phi$,
\be
\D_1\ln \frac{\D_2\Phi+\D_3\Phi}{D_2\Phi-\D_3\Phi}=\D_2\ln
\frac{\D_1\Phi+\D_3\Phi}{\D_1\Phi-\D_3\Phi},
\label{dDSKP}
\ee
and this equation coincides with continuum limit of discrete SBKP equation introduced in
\cite{KS2}.

Equation (\ref{dDSKP}) can be written in symmetric form,
\be
\Phi_{23}\Phi_1(\Phi_2^2-\Phi_3^2)+\Phi_{31}\Phi_2(\Phi_3^2-\Phi_1^2)+\Phi_{12}\Phi_3(\Phi_1^2-\Phi_2^2)=0,
\ee
where subscripts denote vertex derivatives.

It is also easy to find a determinant representation for this equation,
\be
\det
\left(
\begin{array}{ccc}
\Phi_1&\Phi_2&\Phi_3\\
\Phi_2\Phi_3&\Phi_1\Phi_3&\Phi_1\Phi_2\\
\Phi_{23}&\Phi_{13}&\Phi_{12}
\end{array}
\right)
=0.
\label{BKPconfdet}
\ee

We have defined the function $\Phi$ implicitly through the relation $p_i=\D_i\Phi$.
However, it is possible to construct this function explicitly, using the potential
$u$ of dispersionless hierarchy. Indeed, $u=-F_x$, $p(z)=-\partial_x\D(z)F+z$, then
$p(z)=z+\D(z)u$. Thus $\Phi$ is `almost' $u$. For $u$, instead of (\ref{dDSKP}),
we get a generating equation
\be
\D_1\ln \frac{u_2+u_3+z_2-z_3}{u_2-u_3+z_2-z_3}=\D_2\ln \frac{u_1+u_3+z_1+z_3}{u_1-u_3+z_1-z_3}.
\label{dDSKP1}
\ee
This equation gives dBKP hierarchy equations for potential $u$ by expansion
into parameters $z_1, z_2, z_3$.

To transform relation $p(z)=z+\D(z)u$ to the relation
$p_i=\D_i\Phi$, we define $\Phi$ as $u-\sum_{k=0}^N c_{2k+1}t_{2k+1}$.
It is easy to see that it is possible to satisfy relations
$p_i=\D_i\Phi$ at some set of points $z_i$ by the choice of
constants $c_{2k+1}$ (taking sufficiently large $N$).
Indeed,
$$
p(z)-D(z)\Phi=z+\sum_{k=0}^N \frac{c_{2k+1}}{2k+1}z^{-(2k+1)},
$$
and the relation $p_i=\D_i\Phi$ is satisfied if $z_i$ is a zero of polynomial
$$
P(z)=(z^2)^{N+1}+\sum_{k=0}^N \frac{c_{2k+1}}{2k+1}(z^2)^{N-k}.
$$
Taking, e.g., $P(z)=\prod_{i=1}^{I}(z^2-z_i^2)$, it is possible
to express the constants $c_{2k+1}$ through $z_i$ explicitly.
Thus solution to
equation (\ref{dDSKP}) can be expressed in terms of potential $u$.

\subsubsection*{A general conformally invariant equation of dBKP hierarchy}
It is also possible,
using the approach developed in \cite{dtau}, to derive a general equation
for the symmetry $\phi$ of the function $F$ which is invariant under
conformal transformation (we mean that $f(\phi)$ is also a symmetry for
arbitrary analytic $f$). We start with addition formula (\ref{add3}).
Considering a symmetry $\delta F=e^{\Theta \phi}$, where $\Theta$ is an arbitrary parameter,
we get a system of equations
\begin{eqnarray}
\left\{
\begin{array}{l}
x+y+z=-1,\\
(\phi_{12}+\phi_{13})x+(\phi_{23}+\phi_{12})y+(\phi_{31}+\phi_{23})z=0,\\
(\phi_1\phi_2+\phi_1\phi_3)x+(\phi_2\phi_3+\phi_1\phi_2)y+(\phi_3\phi_1+\phi_2\phi_3)z=0,
\end{array}
\right.
\label{system-conf}
\end{eqnarray}
where $x=c_2c_3 e^{-(F_{12}+F_{13})}$, $y=c_1c_3e^{-(F_{23}+F_{12})}$,
$z=c_1c_2 e^{-(F_{31}+F_{23})}$. The first line of system (\ref{system-conf})
(zero order in $\Theta$)
is addition formula (\ref{add3}), the second (first order in $\Theta$)
defines its symmetry, and the third (second order in $\Theta$)
follows from conformal invariance.

Using this system, we express $x,y,z$ through $\phi$, then find $e^{F_{23}}$, $e^{F_{13}}$
in terms of $\phi$ and get a compatibility condition
\bea
&&
\D_1\ln \frac{(f^1_{23}+f^2_{31}+f^3_{12})
(f^1_{32}+f^2_{31}+f^3_{12})}
{(f^1_{23}+f^2_{13}+f^3_{12})
(f^1_{23}+f^2_{31}+f^3_{21})}=
\nn\\
&&\qquad\qquad
\D_2\ln \frac{(f^1_{23}+f^2_{31}+f^3_{12})
(f^1_{23}+f^2_{13}+f^3_{12})}
{(f^1_{23}+f^2_{31}+f^3_{21})
(f^1_{32}+f^2_{31}+f^3_{12})},
\label{gen-conf}
\eea
where $f^i_{jk}=\D_i\ln\frac{\phi_j}{\phi_k}$. Equation (\ref{gen-conf}) is
a general equation for conformally-invariant symmetry of the $\tau$-function
of dBKP hierarchy.
\section{Symmetry constraints for dBKP hierarchy}
Using the symmetry (\ref{sym3}), we define a symmetry constraint
$$
F_x=\sum_{i=1}^N c_i S_i,
$$
or, in terms of potential $u$,
$$
u=
2\sum_{i=1}^N c_i p_i,\quad p_i=p(z_i).
$$
Evaluating first dBKP Hamilton-Jacobi equation
$$
S_y=p^3 + 3up
$$
where $y=t_3$,
at $z$ equal to
$z_i$, we  get a system of hydrodynamic type
\be
\partial_y p_k=\partial_x
\bigl(p_k^3+6p_k\sum_{i} c_i p_i
 \bigr).
\label{Bhydro1}
\ee
Higher Hamilton-Jacobi equations will give higher systems of constrained dBKP hierarchy.
The Sato function $z(p)$ for this hierarchy is constructed similar to constrained dKP case
\cite{dconstr},
\be
z=p-\sum_{i=1}^N c_i
\ln\frac{p-p_i}{p+p_i}.
\label{zBKP}
\ee
Its expansion at infinity is
\be
z\rightarrow p+\sum_{n=0}^{\infty}{v_{2n+1}}p^{-(2n+1)},
\quad v_{2n+1}=\frac{2}{2n+1}\sum_{i=1}^{N}c_ip_i^{2n+1}.
\label{zBKP2a}
\ee
From (\ref{DEB}) we obtain a generating system for the constrained hierarchy,
\be
\D(z) p_k=-\partial_x \ln \frac{p-p_k}{p+p_k},
\label{gen}
\ee
where $p$ is a function of $z$,
$(p_1,\dots,p_N)$,
defined by the relation (\ref{zBKP}). Expanding both sides
of this system into the powers of ${z}^{-1}$, one gets the systems (\ref{Bhydro1})
and its higher counterparts. Expansion of $p(z)$  at infinity is given by the formula
$$
p(z)=z+\sum_{n=0}^\infty \frac{1}{2n+1}\mathrm{res}_{p=\infty}\bigl(z(p)^{2n+1}\bigr)
z^{-(2n+1)}.
$$

In the same manner, it is possible to define constrained hierarchy
using symmetry (\ref{sym2}) (which can be considered as a special case
of (\ref{sym3})) and (\ref{sym1}). We will give the basic formulae for
constrained hierarchy connected with (\ref{sym2}) and obtain constrained
hierarchy for the symmetry of the type (\ref{sym1}) as a limit.

Using the symmetry (\ref{sym2}), we define a symmetry constraint
$$
F_x=\frac{1}{2}\sum_{i=1}^N c_i (S_i-\tilde S_i),
$$
or, in terms of $u$,
$$
u=\sum_{i=1}^N c_i (p_i-\tilde p_i).
$$
The first hydrodynamic type system of constrained hierarchy is
\bea
\left\{
\begin{array}{l}
\displaystyle
\partial_y p_k=\partial_x
\bigl((p_k^3)+3p_k\sum_{i} c_i (p_i-\tilde p_i)
\bigr)
\\
\displaystyle
\partial_y \tilde p_k=\partial_x
\bigl((\tilde p_k^3)+3\tilde p_k\sum_{i} c_i (p_i-\tilde p_i)\bigr)
\end{array}
\right.
\label{hydro2aB}
\eea
The Sato function for constrained hierarchy is given by
\beaa
z=p-{\textstyle\frac{1}{2}}\sum_{i=1}^N c_i
\ln\frac{p-p_i}{p+p_i}\frac{p+\tilde p_i}{p-\tilde p_i},
\eeaa
The generating equation for the constrained hierarchy is
\beaa
\D(z) p_k=-{\textstyle\frac{1}{2}}\partial_x
\ln \frac{p-p_k}{p+p_k}\frac{p+\tilde p_i}{p-\tilde p_i}.
\eeaa

Finally, we will consider symmetry constraint connected with the symmetry
(\ref{sym3}),
$$
F_x=\sum_{i=1}^Nc_i S_z(z_i),\quad u=\sum_{i=1}^Nc_i \phi_i,
\quad \phi_i=\partial_x S_z(z_i).
$$
Though it is possible to consider this constrained hierarchy directly, we will
obtain it as a limit of the previous case, when $p_i\rightarrow \tilde p_i$.
Then from (\ref{hydro2aB}) we obtain a first system of constrained hierarchy,
\bea
\left\{
\begin{array}{l}
\displaystyle
\partial_y p_k=\partial_x
\bigl((p_k^3)+3p_k\sum_{i}c_i\phi_i
\bigr)
\\
\displaystyle
\partial_y \phi_k=\partial_x
\bigl(3p_k^2\phi_k+3\phi_k\sum_{i} c_i \phi_i\bigr)
\end{array}
\right.
\label{hydro2aB2}
\eea
The Sato function for the constrained hierarchy is
\beaa
z=p+\sum_{i=1}^N c_i
\frac{p\phi_i}{p^2-p^2_i}.
\eeaa
\subsection*{Acknowledgments}
LVB was supported in part by RFBR grant
01-01-00929 and President of Russia grant 1716-2003; BGK was supported in part
by the grant COFIN 2002 `Sintesi'.

\end{document}